 \documentclass[final,3p,times]{elsarticle}

\usepackage[english]{babel}

\usepackage{amsmath}
\usepackage{graphicx}
\usepackage[colorlinks=true, allcolors=blue]{hyperref}
\usepackage{xcolor}
\usepackage{ulem}

\DeclareRobustCommand{\vect}[1]{
  \ifcat#1\relax
    \boldsymbol{#1}
  \else
    \mathbf{#1}
  \fi}

  \newcommand{\cbr}[1]{\left(#1\right)}
  \newcommand{\sbr}[1]{\left[#1\right]}

  \newcommand{\marco}[1]{#1}
  \newcommand{\alb}[1]{#1}
  
\begin{document}

\title{Control of friction: shortcuts and optimization for the rate- and state-variable equation}

 \author[label1]{Andrea Plati}
 \affiliation[label1]{organization={Université Paris-Saclay CNRS Laboratoire de Physique des Solides},
             addressline={1 rue Nicolas Appert},
             city={Orsay},
             postcode={91405},
             country={France}}
\author[label2,label3]{Alberto Petri}
\affiliation[label2]{organization={CNR, Istituto dei Sistemi Complessi},
             addressline={Università Sapienza, P.le A. Moro 5},
             city={Rome},
             postcode={00185},
             country={Italy}}
\affiliation[label3]{organization={Centro Ricerche Enrico Fermi},
             addressline={via Panisperna 89A},
             city={Rome},
             postcode={00184},
             country={Italy}}             
\author[label2]{Marco Baldovin}


\begin{abstract}

 Frictional forces are a key ingredient of any physical description of the macroscopic world, as they account for the phenomena causing transformation of mechanical energy into heat. They are ubiquitous in nature, and a wide range of practical applications involve the manipulation of physical systems where friction plays a crucial role. 
 In this paper, we apply control theory to dynamics governed by the paradigmatic rate- and state-variable law for solid-on-solid friction. Several control problems are considered for the case of a slider dragged on a surface by an elastic spring. By using swift state-to-state protocols, we show how to drive the system between two arbitrary stationary states characterized by different constant sliding velocities in a given  time. Remarkably, this task proves to be feasible even when specific constraints are imposed on the dynamics, such as preventing the instantaneous sliding velocity or the frictional force from exceeding a prescribed bound. 
 The derived driving protocols also allow to avoid a stick-slip instability, which instead occurs when velocity is suddenly switched.
 By exploiting variational methods, we also address the functional minimization problem of finding the optimal protocol that connects two steady states in a specified time, while minimizing the work done by the friction. We find that the optimal strategy can change qualitatively depending on the time imposed for the duration of the process. 
 Our results mark a significant step forward in establishing a theoretical framework for control problems in the presence of friction and naturally pave the way for future experiments.
\end{abstract}

\maketitle

\section{Introduction}\label{sec:intro}
Friction between solids is a complex and not yet well understood transformation of mechanical energy into heat, which can involve destructive processes such as wear. Needless to say, it is important not only for practical applications, 
but also for the description and understanding of many physical phenomena occurring at different scales {\cite{Vanossi2013,Baumberger2006}.
It seems that the simplest laws of friction  were first found by Da Vinci  \alb{\cite{Hutchings2016}} and then rediscovered and formalized by Amontons~\alb{\cite{Desplanques2015}}.
Coulomb showed that static and dynamic frictions are different, and understood that the main features of these laws, independence from the area of contact and proportionality to the normal pressure, could be explained by assuming that the actual contact between the solid surfaces only takes place in limited 
regions determined by the surface asperities. 
This approach was developed much more recently by the extensive work of Tabor and Bowden to whom the birth of modern tribology is generally ascribed \cite{Bowden1950}. 

Despite these classical laws are widely taught in school and university classes, their realm of application is not so wide.  Everyone has sometimes experienced that friction decreases when increasing the sliding velocity, a phenomenon that may occur in many different circumstances, and that can as well be interpreted in terms of contacts between asperities. {Indeed, the asperities have less and} less time to interlock \marco{when the sliding velocity increases,} and one body can \marco{therefore} ``surf" upon the other. However, in some cases, further \marco{acceleration} can lead to a point where friction starts to increase again, so the corresponding curve as a function of velocity presents a minimum~\alb{\cite{Heslot1994}}. This behavior, also known as the Stribeck curve for lubricated surfaces, is typical of many industrial situations, but it is not the only possible one: many empirical velocity dependencies have been observed and modelled, which are suitable to be employed in different contexts such as soft matter systems~\cite{Cantat2013,Marchand2020,Plati2022,Plati2024} and  engineering~\cite{Marques2016,Pennestri2016}. 


Dependence of dynamic friction on the relative velocity is not the only feature neglected by the classical laws.  When two bodies remain in static contact for a long period of time, the frictional force between them is observed to increase with the contact age, a fact first noticed by Coulomb \cite{Scholz1998}.   
Extensive experimental laboratory studies on the sliding friction between two solid surfaces led to formulate more general laws. In a series of papers various authors investigated the frictional behavior of many materials, generally at low velocity, \marco{ and realized that the phenomenological behaviour could be efficiently described by the so called ``rate- and state-variable equations'' (R\&S)} (see~\cite{Dieterich1979,Rice1983,Perrin1995}, and e.g.~\cite{Rabinowicz1956,Rabinowicz1992,Scholz1998,Baumberger2006} on the  evolution of friction laws). \marco{This class of laws owes its name to the explicit presence of the velocity (rate variable) and of a memory term (state variable) in the equation for the friction.}
  R\&S laws retain some features of time-independent friction, such as independence of contact area and proportionality to load. 
 {In the formulation of \cite{Dieterich1978,Dieterich1979}, the friction coefficient $\mu$ acting between two solid surfaces sliding on each other depends on the state variable $\theta$ and the sliding velocity $v$ as
 \begin{equation}
 \mu(t) = \mu_0+ a\ln \left(\frac{v(t)}{v_0}\right)+b\ln\left(\frac{\theta(t) v_0}{l}\right)
 \label{RS}
 \end{equation} 
 where $\mu_0$ is the static friction coefficient, $v_0$ is a reference  {positive} velocity, usually very small with respect to the typical velocities of the system (i.e. $v>v_0>0$), and $l$ is a characteristic length related to the asperity spatial size and correlation.  The state variable $\theta$ represents a characteristic lifetime of contact between the asperities of the two solid surfaces. 
 The positive coefficients $a$ and $b$ determine the weight of the rate-dependent and state-dependent terms, respectively. Both of these terms contribute to increasing the friction coefficient in Eq.~(\ref{RS}). As we shall see in the next section, $\theta$ is a time dependent variable whose evolution depends on $v$ and this introduces a non-trivial dynamical interplay between friction and sliding velocity.
R\&S laws have been widely used in the description and simulation of coseismic activity~\cite{Scholz1998,Marone1998,Petrillo2020} and have \marco{also} been shown to reproduce many features of crustal earthquakes \alb{\cite{Bizzarri2014,Bizzarri2019}}}

The knowledge of a reliable law for the friction dynamics paves the way to the solution of many problems of practical interest. One may think, for instance, of the quite common situation in which a load needs to be transported along a rough surface, pulled by a mechanical force. The apparently trivial operation of bringing the system, initially at rest, to a motion at constant speed may actually require some time and effort: indeed, in the presence of high friction, the dynamics is unstable and it typically shows ``stick-slip'' regimes~\alb{\cite{Berman1996,Dieterich1978}}, characterized by the fast alternation of accelerating and decelerating motion. One may then ask how to steer the external force, in such a way that the transition occurs in a given time, or is performed in the most efficient way (e.g., spending the minimal energy).

This kind of problems is the focus of control theory, a discipline at the boundary of  engineering and mathematics~\cite{kirk2004optimal, pontryagin1987mathematical}. The goal is to find protocols to steer controllable systems {between} given initial and  final states, while fulfilling specific conditions. During the last few decades, the application of control theory to physics has seen a renovated interest, also because of the conceptual insight it is able to provide. Examples range from quantum mechanics~\cite{guery2019} to stochastic thermodynamics~\cite{schmiedl2007, MaPeGuTrCi2016, guery2023driving, loos2024universal}, with recent developments involving out-of-equilibrium systems such as granular materials~\cite{PhysRevResearch.3.023128, ruiz2022optimal}, active matter~\cite{shankar2022optimal, baldovin2023, davis2023active}, turbulent fluids~\cite{calascibetta2023optimal} and resetting systems~\cite{de2023resetting, goerlich2024resetting}. 

The aim of this paper is to show how control problems can be addressed in the presence of friction. The simple mechanical model we consider, described in Sec.~\ref{sec:model}, is composed by a block sliding on a rough surface under the effect of an elastic force. We assume that the dynamical friction coefficient is well described by the R\&S.  In Section~\ref{sec:rigid}, the simpler limit of ``rigid'' spring is considered, where the elastic constant is assumed to be so stiff that the velocity of the block can be basically steered at wish. In this regime, we \alb{ solve} the problem of bridging two stationary states corresponding to different constant velocities. We also consider the case of fulfilling additional constraints such as an upper bound on the friction force, or the minimization of the mechanical work. The latter is analyzed in some detail, due to its relevance from the point of view of both theoretical physics and practical applications. Section~\ref{sec:nonrigid} is devoted to the more general case of a non-rigid spring, whose solution can be built exploiting the results found previously. The case of a smooth connection between steady states is particularly relevant in this case, because it avoids the instability towards the stick-slip regime that would be encountered with an abrupt change of the driving velocity. Finally, in Section~\ref{sec:conclusions} we draw our conclusions.

\section{Model}
\label{sec:model}
Let us consider a body of mass $M$ sliding over a rough horizontal surface, pulled by a spring whose other extremity can be externally driven. We limit our analysis to motions occurring along a given direction, say $x$. The considered body is subject to the action of the elastic force
\begin{equation}
    F(t)=k s(t)\,,
\end{equation}
along $x$, where $k$ and $s$ represent the stiffness and the elongation of the spring, respectively. Since the body is in contact with a rough surface, a friction force $\tau$ opposing to the motion is also present. Calling $v$ the velocity of the body, one has therefore
\begin{equation}
\label{eq:dyn}
    m\dot{v}(t)=ks(t)-\tau(t)\,.
\end{equation}
The elongation $s$ obeys in turn the simple dynamical equation
\begin{equation}
\label{eq:dynL}
    \dot{s}(t)=v_c(t)-v(t)\,,
\end{equation}
where $v_c$ is the velocity of the end of the spring that is externally controlled. The situation is sketched in Fig.~\ref{fig:sketch}.
\begin{figure}
    \centering    \includegraphics[width=0.6\linewidth]{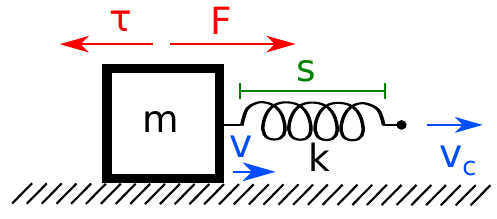}
    \caption{Sketch of the system. The mass $m$ is pulled by the force $F$ due to the action of a spring of stiffness $k$, whose instantaneous elongation is denoted by $s$. The speed of the of the externally controlled end of the spring is denoted as $v_c$, while that of the block is denoted as $v$. The friction force $\tau$ is exerted by the rough ground.}
    \label{fig:sketch}
\end{figure}

We assume the friction force $\tau$ to depend on $v$ according to the phenomenological ``rate- and state-variable equation'' already introduced in the Introduction~ {~\cite{Dieterich1978,Dieterich1979,Baumberger1999}}:
\begin{equation}
\label{eq:ras1}
   \tau(t)= \sbr{\mu_0+a \ln\cbr{\frac{v(t)}{v_0}}+b\ln \cbr{\frac{v_0\theta(t)}{l}}}mg\,.
\end{equation}
Here $g$ is the gravity acceleration  {and the expression in the square brackets coincides with Eq.~\eqref{RS} of the introduction.} When the two positive phenomenological parameters $a$ and $b$ are much smaller than the dynamic friction coefficient $\mu_0$, the familiar constant proportionality law between $\tau$ and the normal force is recovered. As soon as $a$ is non-negligible, an explicit dependence on the body's velocity appears, and the friction is increased by a contribution that is logarithmic in $v$. This modelization turns out to be accurate for values of $v$ larger than a threshold $v_0$, to be determined experimentally{~\cite{Baumberger1999}}. Finally, if $b$ is also \alb{not small,  the} additional term that depends on the {state variable} $\theta$ must be taken into account.
%
This quantity obeys the evolution equation
\begin{equation}
\label{eq:ras2}
\dot{\theta}(t)= 1-\frac{\theta(t) v(t)}{l}\,,
\end{equation}
where the typical length $l$ is also phenomenological. The presence of $\theta$ in Eq.~\eqref{eq:ras1} induces memory effects on the friction evolution, whose time-scale is of order $l/v$. {This is consistent with the idea introduced in Sec.~\ref{sec:intro} that $\theta$ represents the typical contact lifetime between surface asperities. Indeed, Eq.~\eqref{eq:ras2} expresses the fact that the interpenetration of asperities, and consequently their contact lifetime, grows faster at lower velocities.} 
As we shall see, {the memory effects introduced by the dynamics of the state variable} make the search for control protocols non trivial. The situation is qualitatively similar to that of inertial systems in stochastic thermodynamics~\cite{PMG2014,baldovin2022,sanders2024optimal}, where the control of position fluctuations is also influenced by the instantaneous momentum of the particles -- making the problem considerably more difficult than its overdamped counterpart.

The considered dynamics admits a stationary state where $v$, $v_c$, $\theta$ and $\tau$ are constant. In this regime,
\begin{subequations}
\label{eq:ststate}
\begin{eqnarray}
\label{eq:vst}
v&=&v_c  \\
\label{eq:thetast}
\theta&=&\frac{l}{v_c}  \\
\label{eq:taust}
\tau&=&\sbr{\mu_0+(a-b)\ln \cbr{\frac{v_c}{v_0}}}mg \,. 
\end{eqnarray}
\end{subequations}
This state is attained through an exponential relaxation after $v_c$ is set to a constant value.

The stability of these stationary states {with respect to a stick-slip dynamical regime} has been thoroughly studied in{~\cite{Gu1991}}, as a function of the system's parameters (see also~\ref{sec:stickslip}). It is worth noticing that if $a>b$, the stationary friction force increases with $v_c$, hence the denomination ``velocity strengthening''. Similarly, ``velocity weakening'' refers to the case $a<b$.

In the rest of the paper, we will describe control protocols to be applied to $v_c$, in order to complete preassigned transitions between stationary states, under given conditions (bounds, optimality). The considered examples are meant to illustrate a quite general technique that might be applied to different problems of practical interest, while underlining the intrinsic difficulties due to the presence of memory effects.

\section{Control in the limit of stiff spring}
\label{sec:rigid}

In this section we restrict our attention to the limit of stiff spring, where the elongation $s$ is almost constant and
\begin{equation}
    v \simeq v_c\,.
\end{equation}
In order to verify this condition, the spring must be stiff enough for the oscillation amplitude to be negligible with respect to a typical lengthscale (e.g., the distance covered by the body during the protocol). For instance, in the case of constant driving velocity $v_c$ the request is fulfilled whenever 
$$
v_c \sqrt{m/k} \ll v_c t_f\,,
$$
or, equivalently,
\begin{equation}
\label{eq:kapprox}
k \gg \frac{m}{t_f^2}\,.
\end{equation}
Controlling $v_c$ is thus equivalent to controlling the velocity $v$ of the mass itself. Extensions to the case of non-stiff spring will be addressed in Sec.~\ref{sec:nonrigid}.

To simplify the mathematical description of the problem, from now on we will make use of the dimensionless variables 
\begin{subequations}
    \begin{eqnarray}
 t'&=&tv_0/l        \\
 v'&=& v/v_0\\
\theta'&=& \theta v_0/l \\
\tau'&=&\tau/(\mu_0mg)
    \end{eqnarray}
\end{subequations}
and constants
\begin{subequations}
    \begin{eqnarray}
 a'&=&a/\mu_0\\
 b'&=&b/\mu_0\,.
    \end{eqnarray}
\end{subequations}
We will omit the prime symbols from now on, with the convention that all considered quantities are dimensionless, unless specified otherwise. Equations~\eqref{eq:ras1}-\eqref{eq:ras2} read then: 
\begin{subequations}
\label{eq:rasRed}
\begin{eqnarray}
\label{eq:rasRed1}
   \tau&=& 1+a \ln v+b\ln\theta\\
\label{eq:rasRed2}
\dot{\theta}&=& 1-\theta v\,.
\end{eqnarray}
\end{subequations}
\subsection{Swift state-to-state protocols}\label{sec:swift}

The first problem we address concerns the transition between two pre-assigned steady states, in a given time $t_f$. In the considered limit, the state of the system is entirely specified by $\theta$ and $v$: the problem reduces therefore to finding a time-dependent protocol $v(t)$ such that
\begin{equation}
\label{eq:bcv}
v(0)=v_i\,,\quad\quad v(t_f)=v_f\,,    
\end{equation}
while, recalling Eq.~\eqref{eq:thetast},
\begin{equation}
\label{eq:bctheta}
\theta(0)=\frac{1}{v_i}\,,\quad\quad \theta(t_f)=\frac{1}{v_f}\,.    
\end{equation}
Requiring that the target state is reached in a given finite time $t_f$ makes the problem non-trivial. Indeed, in order to get to the envisaged final state one could just change the value of $v$ abruptly from $v_i$ to $v_f$, and wait until the new stationary state is reached: however, this relaxation process would take an infinite time~\footnote{In more physical terms, a time much larger than $1/v_f$ in dimensionless units, see Eq.~\eqref{eq:rasRed2}}. The protocols we search for can be thus seen as ``shortcuts'' to the final stationary state (often called ``shortcuts to adiabaticity'' in the literature~\cite{guery2019}, with reference to an analogous problem in quantum mechanics).

In actual experiments instantaneous velocity changes are often difficult to obtain: we will thus limit our attention to protocols $v(t)$ that are continuous in time (hence the denomination ``swift state-to-state'' for this kind of transitions~\cite{guery2023driving}).
The fundamental idea is to search for a possible state evolution $\theta(t)$ among a sufficiently wide family of time-dependent functions 
$$
\theta(t)=f(t;\vect{\alpha})\,,
$$
where $\vect{\alpha}$ is a vector of free parameters to be chosen according to the boundary conditions. In the present case, since we aim to impose the four boundary conditions~\eqref{eq:bcv} and~\eqref{eq:bctheta}, we need $f$ to depend on at least 4 free parameters $\alpha_1, \dots, \alpha_4$.
In the following we will consider some examples, corresponding to different physical constraints.

\subsubsection{Polynomial state evolution}
 A possible choice is the generic third-order polynomial
\begin{equation}\label{eq:polyshort}
    \theta(t)=\sum_{n=1}^4 \alpha_{n}t^{n-1}\,,
\end{equation}
where $\alpha_1, \dots, \alpha_4$ are \alb{set} in such a way that Eqs.~\eqref{eq:bcv} and~\eqref{eq:bctheta} are satisfied. The condition for $\theta(0)$ immediately implies
$$
\alpha_1=1/v_i\,.
$$
Similarly, recalling~\eqref{eq:rasRed2}, the first of Eqs.~\eqref{eq:bcv} yields
$$
\dot{\theta}(0)=1-\frac{1}{v_i}v_i=0
$$
hence
$$
\alpha_2=0\,.
$$
Reasoning in the same way for the conditions at $t=t_f$, one gets
\begin{equation}
\label{eq:poly}
    \theta(t)=\frac{ v_f -3 \,\Delta v\,  (t/t_f)^2 +2  \,\Delta v\,(t/t_f)^3}{v_i v_f  }\,,
\end{equation}
where
$$
\,\Delta v\, = v_f- v_i\,.
$$
By making use again of Eq.~\eqref{eq:rasRed2}, we also obtain
\begin{equation}
   v(t)= \frac{1}{t_f}\frac{ v_i v_f t_f +6  \,\Delta v\,(t/t_f) - 6  \,\Delta v\, (t/t_f)^2}{ v_f -3  \,\Delta v\, (t/t_f)^2+ 2 
   \,\Delta v\,(t/t_f)^3}\,,
\end{equation}
and $\tau(t)$ can be then computed through~\eqref{eq:rasRed1}. In the limit $t_f \to \infty$, the relation $\theta v =1$ holds at all times and the protocol consists of a sequence of stationary states (quasi-stationary limit). If $t_f$ is instead very small, the velocity satisfies the scaling
\begin{equation}
\label{eq:scaling}
v(t) \simeq \frac{1}{t_f}g(t/t_f)\,,
\end{equation}
with
\begin{equation}
\label{eq:gx}
g(x)= \frac{6 \Delta v(x-x^2)}{v_f +\Delta v (2x^3-3x^2)}\,.    
\end{equation}
\begin{figure}
    \centering
    \includegraphics[width=.7\linewidth]{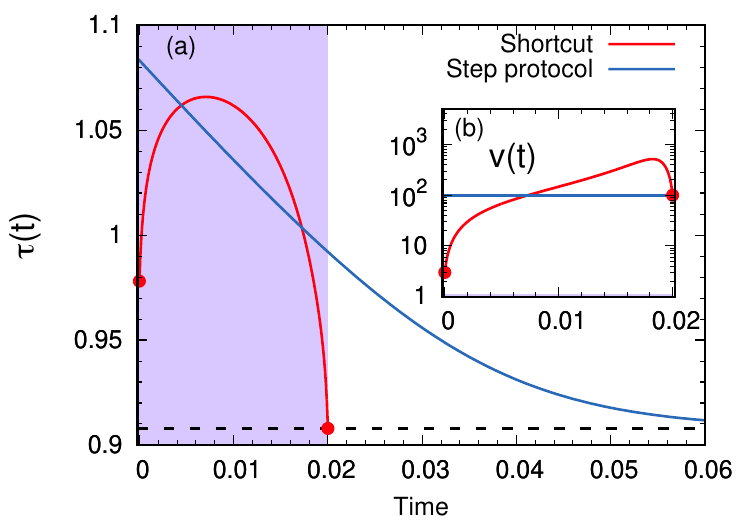}
    \caption{Swift state-to-state transition obtained through the polynomial shortcut~\eqref{eq:poly}. The system starts from a steady state at $v=v_i=3$ and reaches a final steady state at $v=v_f=100$ in a time $t_f=0.02$ (shaded region in the main plot). Panel (a) shows the time-dependence of the friction force $\tau$ along the process (red curve), while the inset (b) represents the corresponding control protocol $v(t)$. For comparison, the same quantities are also plotted in the step-protocol case, where $v$ is abruptly switched from $v_i$ to $v_f$ at time $t=0$. The dashed horizontal line represents the value of $\tau$ in the chosen final state. Parameters: $a=0.03$, $b=0.05$.}
    \label{fig:S1}
\end{figure}
In Fig.~\ref{fig:S1} the protocol just described is plotted for the case of a fast acceleration, in a model with realistic parameters {chosen according to the experimental measurements reported in \cite{Baumberger1999}}. The figure shows that the final target value of $\tau$ can be reached in a time interval fairly shorter than the natural relaxation time of the system. In other words, instantaneously switching the velocity to the target $v_f$ is not the fastest way to get to the desired final state. {As shown in the inset,} the price to pay is that $v(t)$ assumes values much larger than $v_f$ during the transition: this behaviour could also be deduced from the scaling law~\eqref{eq:scaling}.

\subsubsection{Lower bound on the velocity}
\label{sec:lbv}

In the above discussion on the polynomial shortcut, we have disregarded a crucial validity condition for the considered model, namely that the velocity of the body must be larger than the threshold $v_0$. In the considered dimensionless variables, this requirement reads
\begin{equation}
\label{eq:vg1}
    v>1\,.
\end{equation}
Let us consider the case $v_f\ll v_i$ (strong deceleration), and let us focus on the intermediate times of the protocol, $0<t<t_f$. A closer look at Eq.~\eqref{eq:gx} reveals that not only {could} $v(t)$ assume values smaller than 1, but even negative ones.
To enforce the physical condition~\eqref{eq:vg1} we need therefore to choose a different family of functions, such that $v(t)$ be larger than 1 independently of the boundary conditions.

A possible way of enforcing condition~\eqref{eq:vg1} is to set
\begin{equation}
\label{eq:protv}
v(t)=1+\exp\sbr{f(t;\vect{\alpha})}\,,
\end{equation}
where $f(t;\vect{\alpha})$ is, as before, a function of $t$ that also depends on a vector $\vect{\alpha}$ of additional parameters. The velocity is always larger than $1$ by construction. It is important to notice that the state variable $\theta(t)$ is not completely determined by Eq.~\eqref{eq:rasRed2}, once $v(t)$ is chosen: we can still \alb{set} one of the boundary conditions~\eqref{eq:bcv},~\eqref{eq:bctheta}, since the differential equation is first order. 
As a consequence,  in order to enforce the remaining three boundary conditions we only need three parameters $\alpha_1, \alpha_2, \alpha_3$ in the expression of $v(t)$.
By requiring $f$ to be a second-order polynomial and imposing the boundary conditions for $v(t)$, one gets
\begin{equation}
\label{eq:protboundv}
v(t)=1+(v_i-1)\exp\sbr{\ln \left(\frac{v_f-1}{v_i-1}\right)\frac{t}{t_f}-\alpha_3 (t_f-t)t}\,.
\end{equation}
Once the above expression is inserted into~\eqref{eq:rasRed2}, an explicit solution $\theta(t)$ that matches the correct boundary conditions can be found numerically. In Fig.~\ref{fig:S2} an explicit example is shown, for the case of a deceleration process.
\begin{figure}
    \centering
    \includegraphics[width=.7\linewidth]{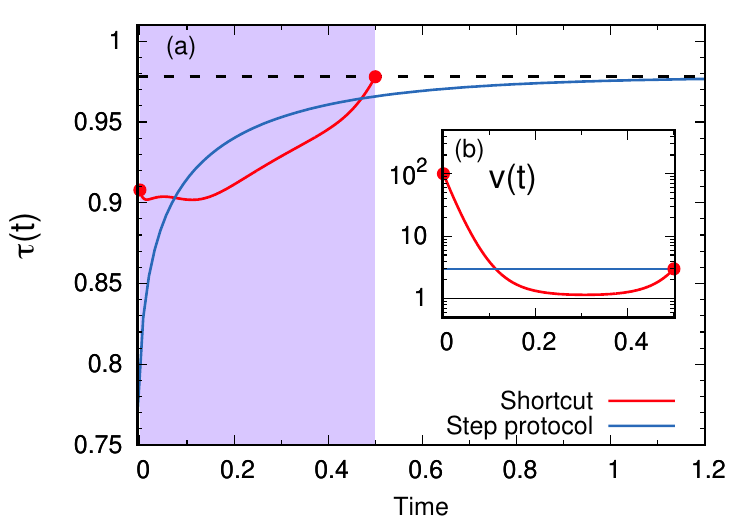}
    \caption{Swift state-to-state transition for a deceleration process, with the condition $v(t)>1$. The protocol belongs to the family~\eqref{eq:protboundv}. In (a) the behaviour of the friction force $\tau(t)$ is shown, while the corresponding control protocol for $v(t)$ is displayed in inset (b): as in Fig.~\ref{fig:S1}, red curves represent the shortcut, while blue ones refer to the step protocol. Here $v_i=100$, $v_f=3$, $t_f=0.5$. Because of the lower bound~\eqref{eq:lbtf}, $t_f$ must be larger than 0.39 in these conditions. Other parameters as in Fig.~\ref{fig:S1}.}
    \label{fig:S2}
\end{figure}

Let us notice that if $v_f<v_i$, and therefore $\theta(t_f)>\theta(0)$, the total time $t_f$ has a trivial lower bound.
Indeed, from Eq.~\eqref{eq:rasRed2} and the condition $v > 1$, one gets that $\theta$ is always positive and then
$$
\dot{\theta}< 1-\theta
$$
or, formally,
$$
dt > \frac{d\theta}{1-\theta}\,.
$$
Integration of both sides of the inequality yields
\begin{equation}
\label{eq:lbtf}
    t_f \ge \ln \cbr{\frac{v_f}{v_i}}+\ln \cbr{\frac{v_i-1}{v_f-1}}\,,
\end{equation}
which is larger than zero in the deceleration case.

\subsubsection{Upper bound on the friction force}
In practical applications, one may require the value of the friction force not to exceed a given threshold $\tau_M$. Also this kind of condition can be addressed in the swift state-to-state transition framework. Reasoning as before, we introduce a suitable function $f(t;\vect{\alpha})$ such that
\begin{equation}
    \tau(t)=\tau_M-\exp \sbr{f(t;\vect{\alpha})}\,,
\end{equation}
and we \alb{set} two of the three free parameters by enforcing the stationary boundary conditions on $\tau(t)$, namely
$$
\begin{aligned}
\tau(0)&=\tau_i\equiv 1+(a-b)\ln v_i\\
\tau(t_f)&=\tau_f\equiv 1+(a-b)\ln v_f\,.
\end{aligned}
$$
As a result, we get the family of functions
\begin{equation}
    \label{eq:taum}
\tau(t)=\tau _M+\left(\tau_i-\tau_M\right)   \left(\frac{\tau_f+\tau _M-2}{
\tau_i+\tau _M-2}\right)^{t/t_f}e^{\alpha_3 t
   \left(t_f-t\right)/{t_f^2}}\,,
\end{equation} 
where $\alpha_3$ needs to be \alb{determined} by enforcing the boundary conditions~\eqref{eq:bctheta}, a task that can be completed by numerically integrating~\eqref{eq:rasRed2}, with the constraint
$$
v(t)=\exp \sbr{\frac{\tau(t)-b \ln \theta(t) -1}{a}}\,
$$
provided by Eq.~\eqref{eq:rasRed1}. Figure~\ref{fig:S3} provides an explicit example.

\begin{figure}
    \centering
    \includegraphics[width=.7\linewidth]{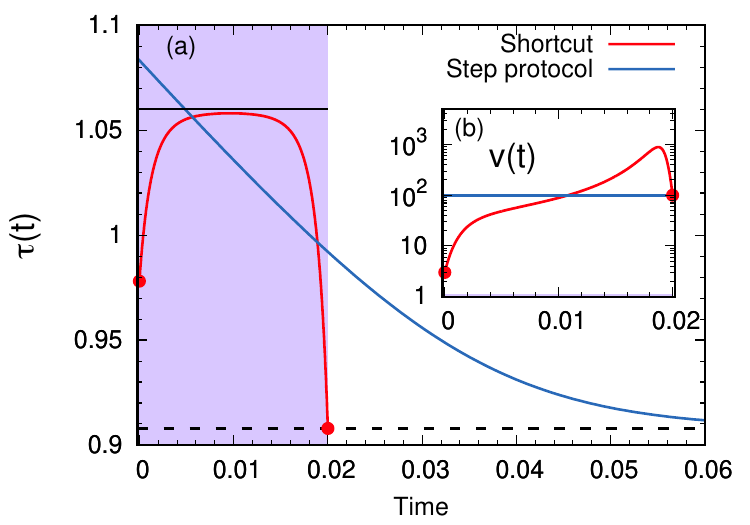}
    \caption{Swift state-to-state transition in the presence of an upper bound $\tau_M$ for the friction force. The boundary conditions are the same as in Fig.~\ref{fig:S1}, but we impose the condition $\tau(t)<\tau_M=1.6$ through Eq.~\eqref{eq:taum}\,. Other details as in Fig.~\ref{fig:S2}.}
    \label{fig:S3}
\end{figure}

\subsection{Optimal control}\label{sec:opt_block}

We now focus on the optimization problem. \marco{Referring again} to the dynamical system defined by Eqs. \eqref{eq:rasRed}, our goal is to find the optimal protocol $v(t)$ to go from \marco{the stationary state at} $v(0)=v_i$ to \marco{the one at} $v(t_f)=v_f$\marco{, in such a way that} the work done by the friction force
\begin{equation}
\label{eq:work}
     W=\int _0^{t_f}\tau(t)v(t)dt 
\end{equation}
is \alb{minimum}. This is a functional minimization problem which can be approached using the formalism of variational calculus by defining the following Lagrangian function  \cite{kirk2004optimal}:
\begin{equation}
    \mathcal{L}(\theta, \dot{\theta},v,\beta)=(1+a \ln v+b\ln\theta)v+\beta(\dot{\theta}-1+\theta v)
\end{equation}
where the first term \marco{on the right hand side} is the argument of the integral we want to minimize, and the second \marco{one} is \marco{a} constraint on the dynamical equation\marco{~\eqref{eq:rasRed2}} which relates $\theta(t)$ and $v(t)$. In order to impose such a constraint, we introduce the Lagrange multiplier $\beta(t)$. \marco{It is important to notice that,} contrary to the standard Lagrange multiplier method for constrained optimization over \marco{finite-dimensional} domains, in this functional optimization problem $\beta(t)$ \marco{is time-dependent}. \marco{We also stress that $v(t)$, our control parameter, can show discontinuities in its time evolution, as it is typical in optimal control problems.} \marco{Lagrangian theory} assures that the dynamical \marco{protocol}  minimizing $W$ while respecting the imposed constraint is defined by
the Euler-Lagrange equations: 
\begin{subequations}
\begin{eqnarray}    
\frac{d}{dt}\left(\frac{\partial\mathcal{L}}{\partial{\dot{\theta}}}\right)-\frac{\partial\mathcal{L}}{\partial{\theta}}&=&0 \label{eq:betadot0}\\
\quad \frac{\partial\mathcal{L}}{\partial{\beta}}&=&0\label{eq:thetadot0}\\
\quad \frac{\partial\mathcal{L}}{\partial{v}}&=&0\label{eq:vdot0}\,.
\end{eqnarray}   
\end{subequations}


Equation~\eqref{eq:betadot0} provides the evolution equation for $\beta(t)$, namely
\begin{equation}
\label{eq:betadot1}
\dot{\beta}=v\left(\frac{b}{\theta}+\beta\right)\,,
\end{equation}
while Eq.~\eqref{eq:thetadot0} is nothing but the dynamical constraint Eq.~\eqref{eq:rasRed2}. The stationarity condition~\eqref{eq:vdot0} reads
\begin{equation}
\label{eq:constr}
1+a \ln v+b\ln\theta+\beta\theta=0\,,    
\end{equation}
which is an algebraic constraint between the considered variables. One possibility to solve the system is to use the above equation to express $v(t)$ as a function of $\theta(t)$ and $\beta(t)$, and then plug this result in~\alb{\eqref{eq:betadot1}} and~\eqref{eq:rasRed2}. We adopt instead a different strategy: we take the time derivative of~\eqref{eq:constr} and use it to find an evolution equation for $v(t)$ as well. While doing so, we have to \alb{set} the initial condition for the evolution of $v(t)$, namely its limit $v(0^+)$, computed for positive $t\to0$ (which can be different from $v_i$, because we admit discontinuous jumps in the control protocol). This additional boundary condition can be deduced from~\eqref{eq:constr} itself, and it reads
$$
v(0^+)=v_i^{b/a}\exp\sbr{-\frac{1}{a}\cbr{\frac{\beta(0)}{v_i}+a+1}}\,.
$$
Summarising, one has to solve the differential system

\begin{subequations}
\label{eq:eqMot}
\begin{eqnarray}
\label{eq:eqMot1}
\dot{\theta}&=&1-\theta v\\
\label{eq:eqMot2}
\dot{\beta}&=&v\left(\frac{b}{\theta}+\beta\right)\\
\label{eq:eqMot3}
\dot{v}&=&-\frac{v}{a}\left(\frac{b}{\theta}+\beta\right)
\end{eqnarray}
\end{subequations}
with the boundary conditions
\begin{subequations}
\label{eq:boundCond}
\begin{eqnarray}
\label{eq:eqMot4}
\theta(0)&=&\frac{1}{v_i}\\
\label{eq:eqMot5}
\theta(t_f)&=&\frac{1}{v_f}\\
\label{eq:eqMot6}
\beta(0)&=&-v_i(1+a+a\ln v(0^+)-b\ln v_i).
\end{eqnarray}
\end{subequations}
This is a well-defined boundary value problem which can be numerically solved using the shooting method \cite{Keller1968} once the numerical values of $v_i$, $v_f$, $t_f$, $a$ and $b$ are specified. 

To test these results, we compare the work $W^*$ done by the friction force within the optimal protocol to that of polynomial shortcuts. As a benchmark, we use the family of fourth-order polynomials Eq.~\eqref{eq:polyshort} for $\theta$ (without imposing the continuity of $v$, for consistency with the optimal protocol). We are therefore left with two free parameters ($\alpha_2$ and $\alpha_3$) to be varied. In this way, we can study how the work $W_s(\alpha_2,\alpha_3)$ done with the shortcuts changes as a function of the parameters. 
\begin{figure}
    \centering
    \includegraphics[width=0.8\linewidth]{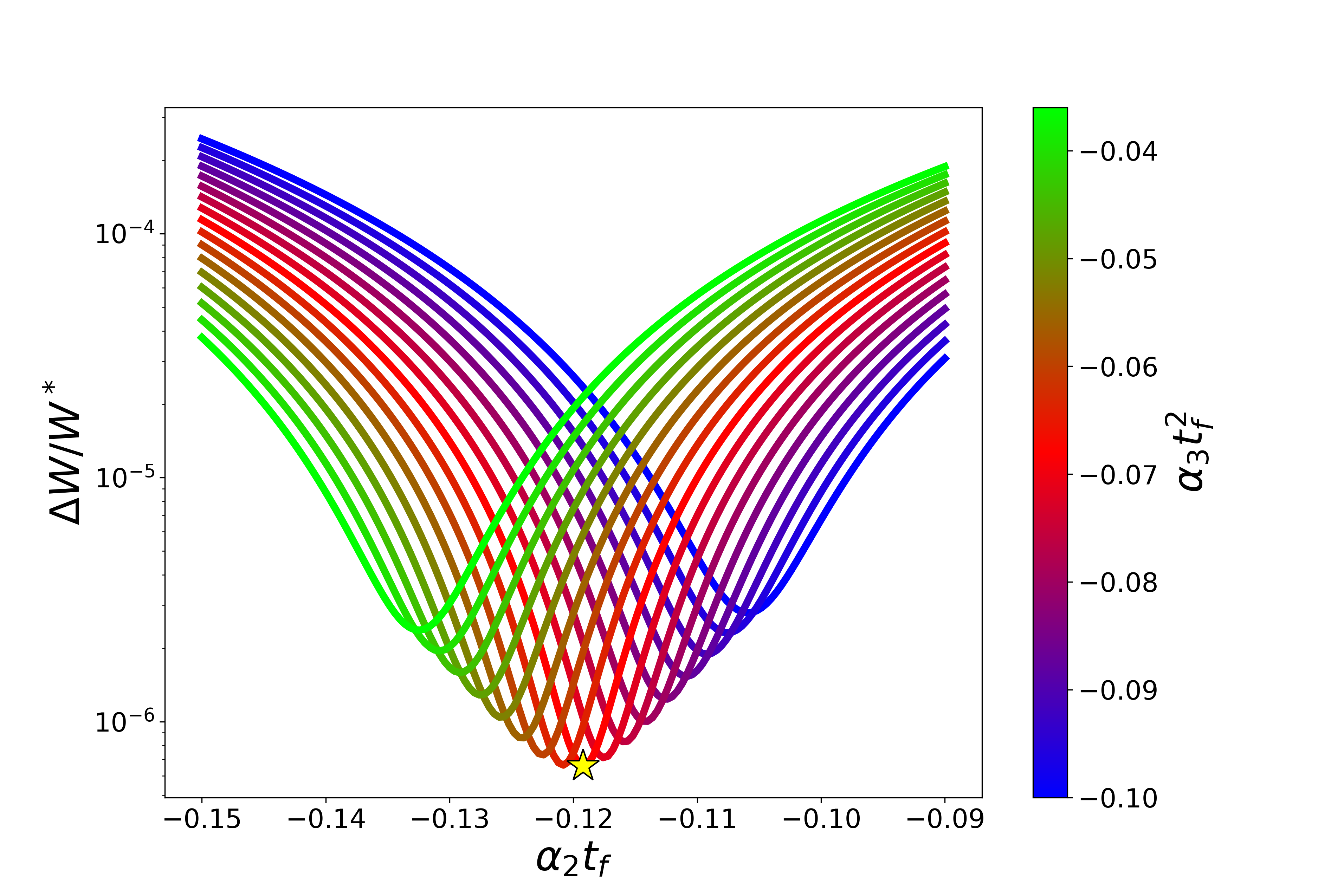}
    \caption{Rescaled difference $\Delta W/W^*=(W_s - W^*)/W^*$ obtained for many combinations of $\alpha_2$ and $\alpha_3$ in the polynomial shortcuts for $\theta$. $W^*$ is the work done by the friction force with the optimal protocol, the yellow star marks the couple of $\alpha_2$ and $\alpha_3$ giving the minimum difference. Numerical values of parameters and boundary conditions are $t_f=0.02$, $a$ = 0.03, $b$ = 0.05,
 $v_i$ = 3 and $v_f$ = 100.}
    \label{fig:compOptSwift02}
\end{figure}
In Fig. \ref{fig:compOptSwift02}, we show the rescaled difference $\Delta W/W^*=(W_s - W^*)/W^*$ for many combinations of $\alpha_2$ and $\alpha_3$. We note that such a difference is always positive, and reaches a minimum very close to zero. This means that \marco{the protocol found with the Lagrangian approach is better than any shortcut protocol of the considered family, as expected; still, }it is possible to find a polynomial shortcut for which the work done by the friction force is approximately the same as the optimal one. The yellow star in the plot marks the \marco{pair} $\alpha_2$ and $\alpha_3$ giving such a ``best shortcut". In Fig. \ref{fig:compOptStep002}, we show the time evolution of $v$, $\theta$ and $\tau$ obtained with the best shortcut and the one obtained with the numerical integration of Eqs. \eqref{eq:eqMot}. As \marco{could be expected} from having found $W_s \sim W^*$, the two sets of curves are \marco{in fact} indistinguishable.
\begin{figure}
    \centering
\includegraphics[width=0.6\linewidth]{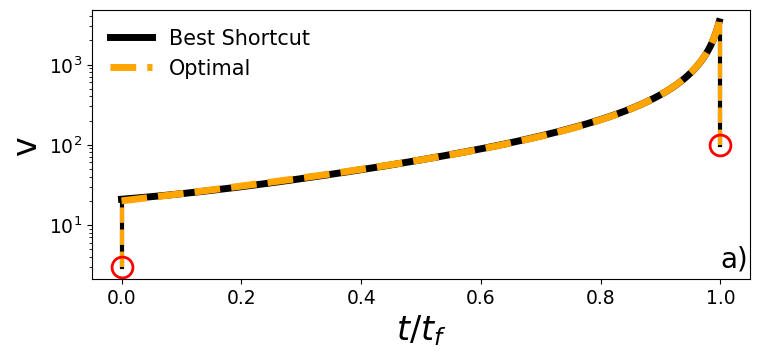}   
\includegraphics[width=0.6\linewidth]{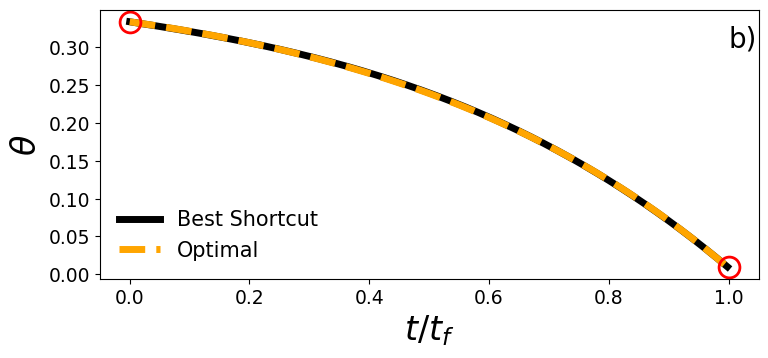}
\includegraphics[width=0.6\linewidth]{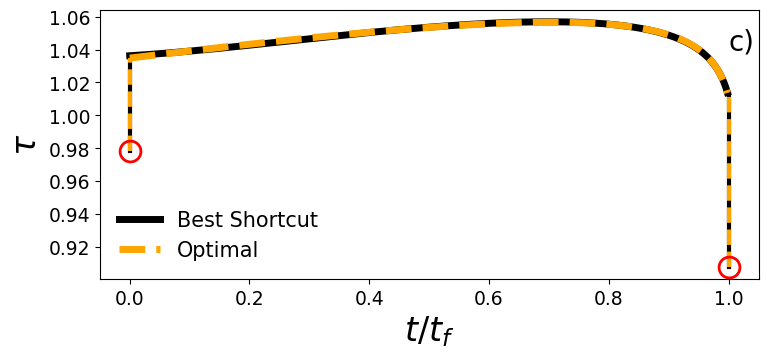}
    \caption{Comparison between the optimal solution and the ``best shortcut" ($\alpha_2=-0.12/t_f$, $\alpha_3=-0.068/t_f^2$) for $v$, $\theta$ and $\tau$. Numerical values of parameters and boundary conditions are $t_f=0.02$, $a$ = 0.03, $b$ = 0.05,
 $v_i$ = 3 and $v_f$ = 100.}
    \label{fig:compOptStep002}
\end{figure}

We note that the optimal protocol consists of a first sharp increase of the velocity followed by a super-exponential growth. Eventually, the velocity is instantaneously switched to $v_f$ from a value which is one order of magnitude larger. During the protocol, $\theta$ is monotonously decreasing and $\tau$ exhibits initial and final jumps (which reflect the discontinuities of $v$) and remains approximately constant in between. 
\marco{We can conclude that} the optimal strategy \marco{prescribes to stay} at a relatively high velocity for the entire protocol duration. This, in turn, promotes the decrease of $\theta$, consistently with Eq.~\eqref{eq:eqMot1}. We recall here that $\theta$ represents {the typical lifetime of the contact} between the {surface} asperities of the two solids and that the friction force Eq.~\eqref{eq:rasRed1} is an increasing function of $\theta$. This tells us that, for the specific imposed time $t_f$, the way to minimize the integral \eqref{eq:work} while satisfying the dynamical constraints is \marco{to reduce} the {contact lifetime} at the cost of having a large velocity during the entire protocol. In Fig. \ref{fig:compVarioTm} we show how the optimal strategy changes for larger $t_f$. Here we plot the optimal $v(t)$, $\theta(t)$ and $\tau(t)$ for three different $t_f$. Blue lines correspond to those shown in Fig. \ref{fig:compOptStep002}, the red and the green ones are obtained with larger $t_f$. Focusing on the \marco{beginning of the protocol for} $v(t)$ and $\theta(t)$, we can identify a crossover between a low-$t_f$ strategy and a high-$t_f$ one.
As already discussed, for $t_f=0.02$, the velocity abruptly increases at the beginning and $\theta$ decreases consequently. On the contrary, for $t_f=0.049$, the protocol starts with a sharp decrease from $v_i$ to $v(t)\sim 1$ (i.e. minimum value allowed for the model's validity). The initial jump is then followed by a super-exponential growth analogous to the one obtained at lower $t_f$. This behaviour of the velocity is translated into a non-monotonic evolution of $\theta(t)$, which is slightly increasing at the beginning, before decreasing towards $1/v_f$. Thus, for large $t_f$, the optimal strategy at short times favours the reduction of the velocity rather than the reduction of the {contact lifetime of the asperities}.
We also \marco{analyzed} an intermediate time $t_f=0.037$ where $v$ is continuous in $t=0$, resulting in a zero initial derivative of $\theta$. Finally, the behaviour of $\tau$ reflects the one of $v$ for the three considered cases. Such a phenomenology is consistent with the two limiting cases we can expect for $t_f \to 0$ and $t_f\to \infty$. In the \marco{former} limit, the only way for the protocol to match the boundary conditions is to jump to very high velocities such that $\theta$ relaxes to $1/v_f$ in the short time $t_f$ and finally switches $v$ to $v_f$. In the \marco{latter}, we can expect that the optimal strategy is to not move the block until an optimal time $t^*$\marco{, quite close to $t_f$,} and then start the protocol to reach the final state. In this way, the work done by the friction force in the time interval $[0,t^*]$ will be zero by definition. 
\begin{figure}
    \centering
\includegraphics[width=0.6\linewidth]{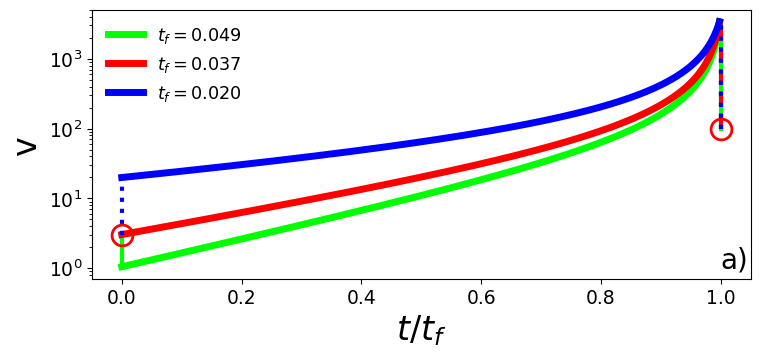}   
\includegraphics[width=0.6\linewidth]{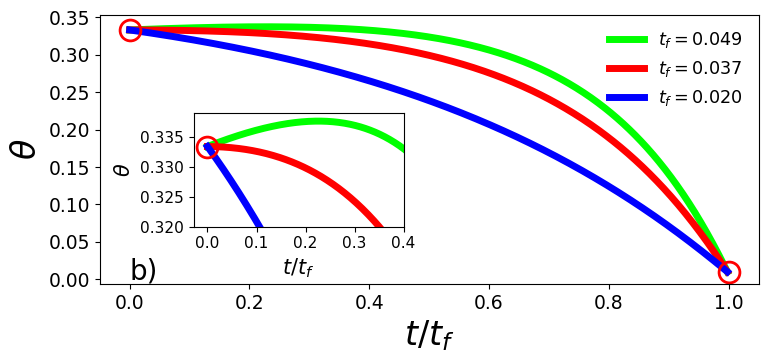}
\includegraphics[width=0.6\linewidth]{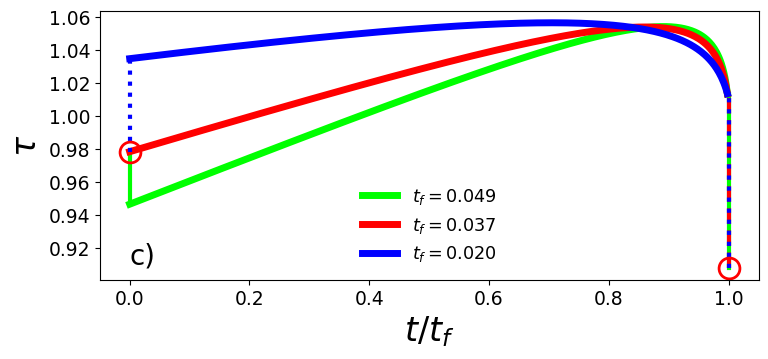}
    \caption{Comparison between optimal solutions obtained for $t_f=0.02,0.037,0.049$. The inset in panel b shows the curves for a different range of values to highlight the non-monotonicity of $\theta$ for $t_f=0.049$. Numerical values of parameters and boundary conditions are $a$ = 0.03, $b$ = 0.05,
 $v_i$ = 3, $v_f$ = 100.}
    \label{fig:compVarioTm}
\end{figure}

By exploring boundary conditions with $v_f>v_i$  {with a larger $t_f$  {or} with $v_f<v_i$,} we observed that the solutions of Eqs.~\eqref{eq:eqMot} \marco{can result in} values of $v(t)$ smaller than $1$, which are not compatible with \marco{our assumptions}.
The way out proposed for the shortcuts in Sec. \ref{sec:swift}, namely defining an auxiliary variable $f$ and taking $v=1+e^f$, is in principle applicable also for the optimization problem but, practically, it brings some non-trivial consequences: indeed, the optimal protocol is formed by a concatenation of extremal and regular values of $v$, a situation that requires more advanced variational methods to be approached~\cite{prados2021, patron2022thermal}. We look forward to applying these techniques to friction models in future works.

\section{Extension to the non-stiff regime}
\label{sec:nonrigid}

In Sec.~\ref{sec:rigid} we computed control protocols for model~\eqref{eq:dyn}, assuming that the spring is so stiff that its elongation $s$ is practically constant and the velocity $v$ of the block is, in fact, directly controlled. Let us now turn to the more general case where $k$ is finite, and the evolution of the spring elongation $s$ also needs to be considered.

By taking the time derivative of Eq.~\eqref{eq:dyn} and plugging Eq.~\eqref{eq:dynL} in it, one obtains
\begin{equation}
\label{eq:vc}
    v_c(t)=v(t)+\frac{\dot{\tau}(t)+m \ddot{v}(t)}{k}\,.
\end{equation}
This relation allows to compute $v_c(t)$, once the process $v(t)$ is known.

\subsection{Swift state-to-state protocols}
Remarkably, shortcut protocols can be found with the same strategy we employed in Sec.~\ref{sec:swift}. It is sufficient to solve the problem for $v(t)$, and then compute the corresponding $v_c(t)$ by mean of Eq.~\eqref{eq:vc}. The only additional difficulty is represented by the fact that $v_c(t)$ depends on $\dot{v}(t)$ (through $\dot{\tau}$) and on $\ddot{v}(t)$: hence, requiring the continuity of $v_c(t)$ at the initial and final times implies the additional boundary conditions
\begin{equation}
\label{eq:bcv1}
\dot{v}(0)=0\,,\quad\quad \dot{v}(t_f)=0\,,    
\end{equation}
and
\begin{equation}
\label{eq:bcv2}
\ddot{v}(0)=0\,,\quad\quad \ddot{v}(t_f)=0\,.
\end{equation}
These conditions impose additional constraints, which need to be fulfilled by increasing by 4 the degree of the polynomial ansatz $f(t,\vect{\alpha})$ in Eq.~\eqref{eq:protv}. The resulting protocol can be found numerically by mean of standard shooting methods.

An example is shown in Fig.~\ref{fig:S5}, where the numerical values of the parameters are taken from Fig.~2 of~\cite{Im2019}. The considered case falls in the class of physical conditions allowing the stick-and-slip phenomenon mentioned in the Introduction, see~\ref{sec:stickslip} for details. As it is clear from  Fig.~\ref{fig:S5}(c) and~(d), if one tries to reach a stationary state by suddenly changing the control spring velocity {$v_c$}, the system undergoes an unstable dynamics that reaches the target state only after a long relaxation time. Moreover, the transient evolution is characterised by a block velocity with peaks that largely overcome $v_f$ (by a factor 6 in the considered example). With our strategy, instead, the desired final state is reached in an arbitrarily short time: here we took a value of $t_f$ corresponding to $0.1$~s in the dimensional units of~\cite{Im2019} (where the step dynamics is explicitly shown to be still far from the relaxed state after $2$~s). As per the $v(t)$, it spans within a much narrower range. Let us notice that, in this specific example, the control velocity $v_c$ becomes negative: this is not in contrast with our original assumptions, since we only required $v(t)$ to be always larger than 1.

Of course, from a practical point of view, the high instability of the dynamics can still be challenging, even if a swift state-to-state protocol is known. To show this point, in Fig.~\ref{fig:S5} we show how the quality of the solution depends on the precision of the control protocol, here represented by the time-step $dt$ used for the numerical integration of the dynamics. If $dt$ is too large, the final state is not precisely reached, and oscillations around the desired target state persist after $t_f$. To actually implement control protocols in these regimes one needs high precision {in their realization}.

\begin{figure}
    \centering
    \includegraphics[width=.6\linewidth]{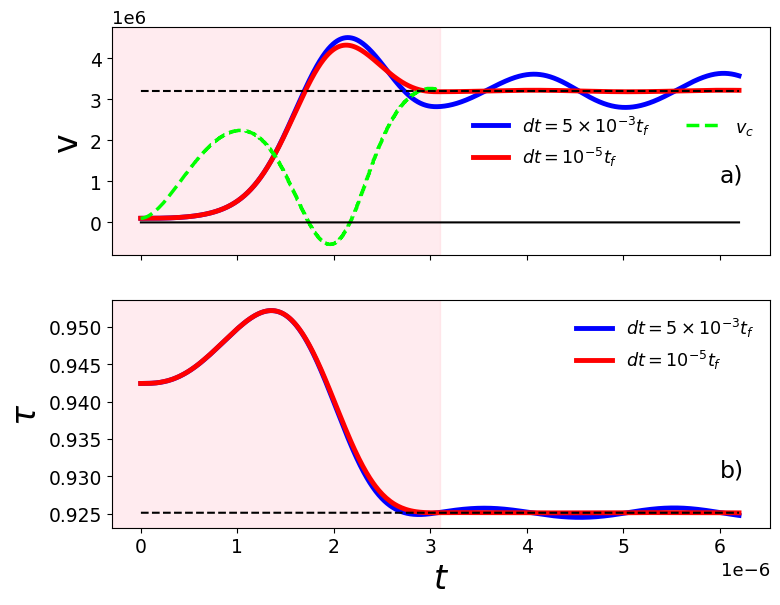}
    \includegraphics[width=.6\linewidth]{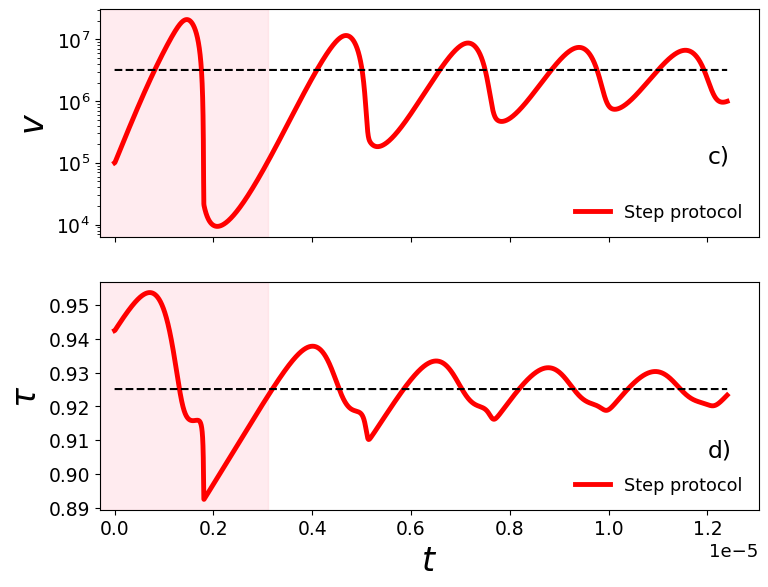}
    \caption{Acceleration process in the non-stiff regime. In panels (a) and (b) the evolution of the block's velocity and friction is shown for the shortcut protocol described in the text, while in panels (c) and (d) the same quantities are reported for the step-protocol where at time $t=0$ the spring velocity is suddenly brought to $v_c=v_f$. The shaded region highlights a time interval $t_f$. For the shortcut protocol, we also show the control velocity $v_c$ (dashed green curve). The shortcut protocol is computed for two different values of the integration step $dt$ (red and blue lines), in order to underline the dependence on control precision in this unstable dynamics. In each plot, dashed black lines signal the target value of the considered quantity. The parameters are inspired by those in Fig.~2 in Ref.~\cite{Im2019}, namely (in dimensionless units): $a=0.005$, $b=0.01$, $m=2.5\times 10^{-16}$, $k=7.5 \times 10^{-3}$, $v_i=10^5$, $v_f=3.2 \times 10^6$, $t_f=3.1\times 10^{-6}$, see~\ref{sec:stickslip} for further details. }
    \label{fig:S5}
\end{figure}

\subsection{Optimal protocols}

Let us now consider the optimal problem discussed in Sec.~\ref{sec:opt_block}. We notice that  the cost function~\eqref{eq:work}, namely the work performed by the friction force, \textit{does not depend on $v_c$}. Therefore we can apply the same line of reasoning used before: the problem of optimizing the work in the case of finite stiffness reduces to the one of controlling $v_c(t)$ in such a way that $v(t)$ is the same as in the infinite-stiffness limit. Therefore, in order to get the optimal protocol $v_c(t)$ we need to substitute into Eq.~\eqref{eq:vc} the expressions for the $v(t)$ and $\tau(t)$ that minimise the work, as computed in Sec.~\ref{sec:opt_block}.

We notice that a problem emerges when computing the second derivative of $v(t)$. Since the optimal protocol involves finite discontinuities at time $t=0$ and $t=t_f$, this means that we get infinite discontinuities for $v_c$ at those boundaries. This is patently unphysical. In order to solve this issue, one possibility is to regularise the protocol with a strategy similar to the one adopted in~\cite{baldovin2023}. The idea is to substitute the solution $v(t)$ in the time intervals $[0,\delta t]$ and $[t_f-\delta t, t_f]$ with two sufficiently smooth shortcuts fulfilling the continuity conditions. In this way the $v_c(t)$ computed through Eq.~\eqref{eq:vc} is continuous by construction, and it tends to the ``true'' optimal protocol in the limit $\delta t \to 0$. At a practical level, the value of $\delta t$ has a lower bound \alb{set} by the precision and the operating range of the experimental setup.

\section{Conclusions}
\label{sec:conclusions}

In this paper, we have considered several control problems concerning the dynamics of a block dragged by an elastic spring on a rough solid surface. The frictional force between the contact surfaces is modelled by \alb{a} rate- and state-dependent law. First, we focused on the large stiffness limit, where the control is performed directly on the block's velocity. Here we derived swift state-to-state protocols to drive the system between two arbitrary stationary states, characterised by a constant sliding velocity. \marco{The connection is realized} in \marco{a given} time\marco{, which can be even very short with respect to the time scales of the dynamics}. These results are extended to cases where we constrain the instantaneous sliding velocity or frictional force not to exceed a prescribed \marco{bound}. 
In the same limit, using variational methods, we have found the optimal protocol that minimises the work of friction while moving the system between two stationary states in a given time $t_f$.
 A transition between two qualitatively different classes of optimal strategies is found when the \marco{connection} time is varied: for small $t_f$, the optimal strategy favours high driving velocities, which weaken the frictional force by reducing the contact surface; for large $t_f$, instead, it favours the reduction of work through lower velocities at the cost of increasing the surface contact. We have then generalized the obtained results to the non-stiff regime where the control is performed on the velocity of the spring extremity. This was found to be particularly useful in applying swift state-to-state protocols in the region of parameters \marco{where} a naive velocity switch is followed by stick-slip instabilities, which prevent the target stationary state from being reached. For future works, we plan to apply the methods proposed here to other frictional models and to develop an experimental setup to test our results on realistic systems.
 
\section*{Acknowledgments}
MB was supported by ERC Advanced Grant RG.BIO (Contract No. 785932). 
A. Plati acknowledges funding from the Agence Nationale de la
Recherche (ANR), grant ANR-21-CE06-0039. Some of the simulations were also done with the code available in the Supplemenatry  materials of Ref.~\cite{Im2019}. We are grateful to the authors.

\appendix

\section{Stick-slip instabilities}
\label{sec:stickslip}

The R\&S  friction laws provides criteria for assessing the onset of instabilities in the steady sliding. First results were concerned with the overdamped case, in which inertia can be neglected. In this case motion instability can be observed when the spring is too soft and its constant is lower than \cite{Rice1983,Ruina1983}:
\begin{equation}
\label{eq:odcritic}
k_{od} = \frac{(b-a)\bar{\sigma}}{l},
\end{equation}
{where $\bar{\sigma}$ is the normal load}.
This relation show that instability requires that $a < b$. When the inertia is not negligible unstable sliding occurs when the spring constant is smaller than \cite{Gu1991,Baumberger2006}\footnote{Note that in \cite{Baumberger1999} the critical value of k/$\bar{\sigma}$ decreases slightly with $v_d$ (Fig. 6), while here it increases quadratically with $v_d$.}
\begin{equation}
\label{eq:mcritic}
k_c= \frac{(b-a)\bar{\sigma}}{l} \left(1+ \frac{Mv_d^2}{\bar{\sigma} a l} \right). 
\end{equation}	
At variance with \ref{eq:odcritic}, the critical value of the spring constant depends on the driving velocity. Thus for a \alb{given} spring constant there is also critical velocity $v_c$
above which the sliding becomes unstable:
\begin{equation}
\label{eq:cdriv}
v_c=\sqrt{\frac{\bar{\sigma} a l}{M}}\left( \frac{kl}{(b-a)\bar{\sigma}} -1 \right). 
\end{equation}

 For the compliant swift in Sec. \ref{sec:nonrigid}, 
we have considered the case illustrated in \cite{Im2019} (Table \ref{tab:parameters}), where
for an abrupt velocity step from $v_i=0.1mm/s$ to $v_f=3.2mm/s$, undamped oscillations are observed.
\begin{table}[h]
\centering
\begin{tabular}{|c|c|c|c|c|c|c|c|}
	\hline
	$\mu_0$ & $a$ & $b$ & $l$ (m) & $v_0$ (ms$^{-1}$)  & $M$ (kg) & $\bar{\sigma}$ (N) & $k$ (Nm$^{-1}$)  \\
	\hline
0.6 & 0.003 & 0.006 & $1\cdot 10^{-5}$ &  $1\cdot 10^{-9}$ & $13\cdot 10^{3}$ & $2\cdot 10^{6}$ & $9\cdot 10^{8}$\\
	\hline
\end{tabular}
\caption{Parameters used in the simulation \cite{Im2019}.}
\label{tab:parameters}
\end{table}

 Adopting adimensional units:
 \[
 \begin{split}
 \mu_0 \rightarrow 1, \, a/\mu_0 \to a, \, b/\mu_0 \to b, \, l \to 1, \, v/v_0 \to v,\\ 
 m v_0^2/(\mu_0 \bar{\sigma} l) \to m, \, \bar{\sigma} \to  1, \,  kl/(\mu_0 \bar{\sigma}) \to k  
\end{split}
\]
one obtains the rescaled values in Table \ref{tab:rescpar}.
 \begin{table}[h]
 	\centering
 	\begin{tabular}{|c|c|c|c|c|c|c|c|}
 		\hline
 		$\mu_0$ & $a$ & $b$ & $l$  & $v_0$   & $m$  & $\sigma$  & $k$  \\
 		\hline
 		1 & 0.005 & 0.01 & 1  & 1 & $2.5\cdot10^{-16}$ & $1$ & $7.5  \cdot 10^{-3}$ \\
 		\hline
 	\end{tabular}
 	\caption{Rescaled parameters used in the compliant case.}
 	\label{tab:rescpar}
 \end{table}

\bibliographystyle{plainurl}
\bibliography{biblio}

\end{document}